\documentclass[aps,pra,showpacs,amsmath,amssymb,amsfonts,lengthcheck,twocolumn,longbibliography,superscriptaddress]{revtex4-1}
\usepackage{dsfont}
\usepackage{graphicx} 
\usepackage{graphics}

\usepackage{mathtools}
\usepackage{dcolumn}    
\usepackage{bm} 
\usepackage{graphicx}
\usepackage{amsmath}    
\usepackage{latexsym}
\usepackage{amsfonts}   
\usepackage{amssymb}
\usepackage{array}      
\usepackage{epsfig}
\usepackage{txfonts}
\usepackage{xcolor}
\usepackage[colorlinks=true,linkcolor=blue,urlcolor=blue,citecolor=blue,pdfusetitle]{hyperref}

\usepackage[normalem]{ulem}

\newcommand{\ket}[1]{\left|#1\right\rangle}
\newcommand{\bra}[1]{\langle#1|}
\newcommand{\braket}[2]{\langle#1|#2\rangle}

\usepackage[none]{hyphenat}

\newcommand{\blah}{blah\\blah\\blah\\blah\\blah}
\begin{document}
\title{Efficiency of Feynman's quantum computer}

\author{R. J. Costales}
\thanks{These authors have contributed equally to this work}
\altaffiliation{Presently at the London Centre for Nanotechnology and the Department of Physics and Astronomy, University College London.}
\affiliation{School of Theoretical Physics, Dublin Institute for Advanced Studies, 10 Burlington Road, Dublin D04 C932, Ireland.}

\author{A. Gunning}
\thanks{These authors have contributed equally to this work}
\altaffiliation{Presently at the Department of Physics, University of Oxford.}
\affiliation{School of Theoretical Physics, Dublin Institute for Advanced Studies, 10 Burlington Road, Dublin D04 C932, Ireland.}

\author{T. Dorlas}
\affiliation{School of Theoretical Physics, Dublin Institute for Advanced Studies, 10 Burlington Road, Dublin D04 C932, Ireland.}

\date{\today}

\begin{abstract}

Feynman’s circuit-to-Hamiltonian construction enables the mapping of a quantum circuit to a time-independent Hamiltonian. This model introduces a Hilbert space made from an ancillary clock register tracking the progress of the computation. In this paper, we explore the efficiency, or run-time, of a quantum computer that directly implements the clock system. This relates to the model's probability of computation completion which we investigate at an established optimal time for an arbitrary number of gates $k$. The relationship between the run-time of the model and the number of gates is obtained both numerically and analytically to be $O(k^{5/3})$. In principle, this is significantly more efficient than the well investigated Feynman-Kitaev model of adiabatic quantum computation with a run-time of $O(k^4)$. We address the challenge which stems from the small window that exists to capture the optimal stopping time, after which there are rapid oscillations of decreasing probability amplitude. We establish a relationship for the time difference between the first and second maximum which scales as O($k^{1/3}$).

\end{abstract}

\maketitle

\section{Introduction}

In 1981, Feynman highlighted the physical limitations of classical computers and proposed using quantum computers to simulate quantum systems \cite{Feynman:1984bi}. In searching for a system Hamiltonian that could perform as a computer, he proposed the original clock Hamiltonian for the purposes of circuit model quantum computation. Since then, there has been significant interest in exploiting his scheme. Aharonov et al. used the model to show the equivalence of adiabatic quantum computation and the circuit model \cite{aharonov2005adiabatic}. Kitaev et al. demonstrated the local Hamiltonian problem is QMA-complete for open quantum systems \cite{kitaev2002classical}. The clock model has been more recently studied in several quantum complexity results \cite{hallgren2013local,gosset2016quantum,Caha_2018,breuckmann2014space}, improving gap bounds \cite{Caha_2018}, establishing a relationship with quantum walk Hamiltonians \cite{Caha_2018,gosset2015universal} and extensions to open quantum systems \cite{tempel2014kitaev}. Modifications have also been made to the model to incorporate forward biasing, additional qubits \cite{Caha_2018,PhysRevA.85.032330}, combinations of several clock registers \cite{gosset2016quantum} to improve success probability, a railroad switch gadget for physical implementations \cite{Nagaj_2010,PhysRevA.85.032330,ciani2019hamiltonian,lloyd2016adiabatic} and noise-assisted transport to remove trapping regions in the clock register \cite{de2013noise}. Feynman's original work has also inspired alternative constructions such as the ‘space-time circuit-to-Hamiltonian’ model in which the global clock is replaced by local clocks for each information carrying particle \cite{janzing2007spin,mizel2001energy,mizel2007simple,breuckmann2014space}, as well as models incorporating tensor networks and use of quantum fault tolerance \cite{anshu2024circuit}. The greatest interest in the Feynman-Kitaev model has followed from Aharonov's work regarding adiabatic computation with most research being devoted to improving the minimum spectral gap bounds on this model \cite{aharonov2005adiabatic, Caha_2018,Bausch_2018,gosset2015universal, Dooley_2020,deift2006improved,mizel2007simple,kato1950adiabatic,jansen2007bounds,born1928beweis}. The heavy analysis on this aspect of the adiabatic model has been in an effort to understand it's efficiency (run-time) which was found to be of the order $O(k^4)$ \cite{aharonov2005adiabatic, born1928beweis,jansen2007bounds} where $k$ is the number of unitary operations. 

In this paper, we provide a direct comparison with the adiabatic model by considering the straightforward circuit implementation of Feynman's original clock Hamiltonian. The major contribution of this work is in demonstrating that the efficiency of the clock model is a significant improvement over the adiabatic implementation. 


\subsection{Feynman's Clock Space}
\begin{figure}[t]\label{fig:circuit}
    \centering
    \includegraphics[width=0.9\linewidth]{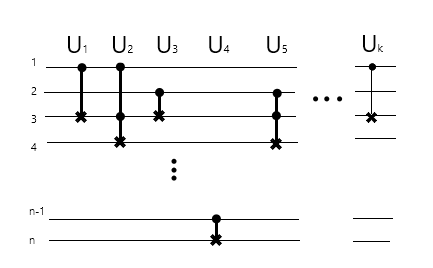}
    \caption{Schematic of a quantum circuit depicting the succession of $\hat{U}_k$ operators we want to perform on the input state, $\ket{\psi_{in}}$, of $n$ qubits in our register to arrive at our desired output state, $\ket{\psi_{out}}$.}
    \label{fig:Operator_k}
\end{figure}

Feynman's clock model exploits the circuit model of quantum computation, originally introduced by Feynman and Deutsch \cite{feynman2018simulating,Deutsch1985QuantumTT}. Here, a calculation is performed by preparing a set of $n$ logical qubits in the computational basis state 
$\ket{\psi_{\rm in}} = \ket{0_1 0_2 ... 0_n}$ and acting on this state with a suitably chosen unitary operator $\hat{U}$ to obtain a final state $\ket{\psi_{\rm out}} = \hat{U}\,\ket{\psi_{\rm in}}$. The unitary operator is constructed in such a way that a measurement of this final state in the computational basis yields the desired result of the computation. Similar to the operation of NOT, AND and OR gates being sufficient to generate any logical operation, it has been proven that one can approximate any given such unitary matrix $\hat{U}$ by a sequence of unitary gates  $\hat{U}_1$, $\hat{U}_2$,$\dots$,$\hat{U}_{k-1}$,$\hat{U}_k$, chosen from a limited set of basic unitary matrices known as a universal set of quantum gates acting individually on a small subset of the logical qubits \cite{deutsch1989quantum, nielsen2010quantum}. Replacing $\hat{U}$ by this sequence of quantum gates, the resulting output state is $\ket{\psi_{\rm out}} = \hat{U}_k \hat{U}_{k-1} \cdots\hat{U}_{2} \hat{U}_1 \ket{\psi_{\rm in}}$. The successive operation of quantum gates can be represented in a quantum circuit diagram as in Fig.\ref{fig:circuit}. Feynman showed that it is also possible to achieve quantum computation by mapping the quantum circuit 
to a continuous-time evolution with a time-independent Hamiltonian. Here, an initial state evolves according to Schrodinger's equation while still implementing a discrete sequence of unitaries \cite{Feynman:1984bi}. There is advantage in this construction as it allows simulation of time-dependent quantum mechanics using a time-independent framework, having no requirement for active driving fields to perform logical operations \cite{aharonov2005adiabatic,mcclean2013feynman,PhysRevA.70.012304}.

Feynman's circuit-to-Hamiltonian construction involves the composition of unitary operators in the following way. Let $\hat{U}_1$,$\hat{U}_2$...$\hat{U}_{k-1}$, $\hat{U}_k$ be the succession of operations we want to operate on the input state of the $n$-qubits in our `register' $\ket{\psi_{\rm in}}$. 
Adjacent to the register, we add an entirely new set of $k + 1$ qubits, called `program counter sites' which live in an ancillary `clock space', separate from the Hilbert space of the register. The purpose of the clock space is to track the progress of the computation. The program counter can be thought of as a particle hopping between sites of a one-dimensional lattice. The states in the clock space are of the form below,
  \begin{gather*}
        \ket{0}_c = \ket{1000...00} \\
        \ket{1}_c= \ket{0100...00} \\
        \vdots \\
        \ket{k-1}_c = \ket{00...0010} \\
        \ket{k}_c = \ket{00...0001}.
    \end{gather*}
Feynman proposed that we initialise our clock space in the state $\ket{0}_c$ at the beginning of our computation. At each step of the computation i.e. as each operator $U_j$ acts on the register of $n$ qubits, the counter moves to the subsequent $\ket{j}_c$ state of our clock space. Eventually, the program counter evolves to the final state of the clock space $\ket{k}_c$ with the final operator $U_k$ acting on the register, at which stage the computation is complete. The register must be observed immediately to ensure that the program counter does not return back down the program line. The complete Hamiltonian, describing the evolution of the register qubits and the clock qubits, is given by,

\begin{align}\label{eq:Hamiltonian}
   \hat{H} &= \sum_{i=0}^{k-1}\hat{q}_{i+1}^{\dag}\hat{q}_i\hat{U}_{i+1} + \text{H.C} \\
   &= \sum_{i=0}^{k-1} \;\left(\ket{i+1}\bra{i}_c \otimes \hat{U}_{i+1} + \ket{i}\bra{i+1}_c \otimes \hat{U}_{i+1}^{\dag}\right).
\end{align}
Here $\hat{q}_i^\dag$ and $\hat{q}_i$ are creation and annihilation operators at the $i$-th clock site, so that $\hat{q}^\dag_{i+1} \hat{q}_i$ moves the counter one site to the right. Note that the Hermitian conjugate term $\hat{q}^\dag_i \hat{q}_{i+1}$ occurring in 
the H.C. term moves the counter one site to the left, so that the clock can also move back in time.

In summary, our Hilbert space for this model $\mathcal{H} =  \mathcal{H}_{\rm clock} \otimes  \mathcal{H}_{\rm register}$ can be broken into two spaces, a clock register whose $k+1$ orthonormal states track the linear progress of the computation and a data register that contains the $n$ computational qubits that our unitary gates act on. During the computation, we will evolve some initial state $\ket{\psi_{0}} = \ket{0}_c \otimes \ket{\psi_{\rm in}}$ with our clock Hamiltonian and we will arrive at some resulting state of the form,
\begin{equation}
    \ket{\psi_t}=\ket{t}_c \otimes (\hat{U}
_t,...,\hat{U}_1 \ket{\psi_{\rm in}}), t = 0,1,...,k,
\end{equation}
where $\ket{\psi_k} = \ket{k}_c \otimes \ket{\psi_{\rm out}}$ is our final state that contains the answer to our computation \cite{Caha_2018}. 

In this paper, we aim to examine the efficiency of Feynman's quantum computer. For our purposes, this is defined to be the run-time or the time of computation completion for our model. We will work this out by examining the probability $P_k(t)$ that the initial state $\ket{\psi_{0}}$ of our computation evolves to the final state $\ket{\psi_k}$ at time $t$.

\subsection{Adiabatic Quantum Computation}

The significance of our results for the run-time of Feynman's clock model in Section \ref{sec:level4} becomes evident when we compare with a model that has been more heavily investigated - adiabatic quantum computing. We include this particular model as its efficiency has been investigated in great detail with regards to evolution using a clock space \cite{aharonov2005adiabatic,Dooley_2020,deift2006improved,mizel2007simple}. Aharonov et al. \cite{aharonov2005adiabatic} showed the computational equivalence between adiabatic computation and the quantum circuit model, by demonstrating an adiabatic evolution with the same output as that of an arbitrary quantum circuit. The critical element was Feynman's circuit-to-Hamiltonian construction and their method is laid out in Appendix \ref{app: adiabatic}. It was shown by Aharonov \cite{aharonov2005adiabatic} and Kitaev \cite{kitaev2002classical} that the run-time for this model is polynomial in the number of operations $k$ by analysing $H_{\rm clock}(s)$,
\begin{equation} \label{adiabatic}
     \hat{H}_{\rm clock}(s) = \begin{bmatrix}
s/2 & -s/2 & & & & \\
-s/2 & 1 & -s/2 & & & \\
& -s/2 & 1 & -s/2 & & \\
& & \ddots & \ddots & \ddots & \\
& & & -s/2 & 1 & -s/2 \\
& & & & -s/2 & 1 -s/2 \\
\end{bmatrix},
\end{equation}
where $s$ is a rescaled time parameter.

The latter authors estimated that the minimum gap $\Delta$ of $\hat{H}_{\rm clock}$ to be bounded below by $1/(144 k^2)$. This bound was improved by Dooley et al. \cite{Dooley_2020} to 
$\pi^2/(8 k^2) \sim 1.23 k^{-2}$. By the adiabatic theorem
(see Born \cite{born1928beweis}) the time needed for adiabatic
evolution is proportional to $\Delta^{-2}$. This relationship requires that the spectrum is discrete,
and that the eigenvalues do not overlap other than at a few points in time, which is true in this case \cite{Dooley_2020}. (Note that the theorem by Born was generalized to the case of a continuous spectrum by Kato \cite{kato1950adiabatic}. The bound by Jansen \cite{jansen2007bounds} differs from Born and suggests that the time needed for adiabatic evolution is proportional to $\Delta^{-3}$ as they allow for a continuous spectrum. This leads to a run-time of $O(k^6)$. However, the example in Section VI  of \cite{jansen2007bounds} suggests that the quadratic dependence is valid in our case, indeed, the spectrum is obviously discrete.) This leads to  a run-time of the adiabatic algorithm of $O(k^4)$, excluding measurement time. In order to compete with adiabatic quantum computation, the straightforward evolution of Feynman's computer must at the very least be polynomial in run-time and more critically hold a smaller power than $O(k^4)$.

An analysis into the run-time of Feynman's model is carried out thoroughly in Section \ref{sec:level4}. In Section \ref{sub:ham}, we show that Feynman's Hamiltonian can be reduced to a simpler clock Hamiltonian similar to the adiabatic Hamiltonian in (\ref{adiabatic}). We demonstrate that the computation is complete with probability $P_k(\tau) = O(k^{-2/3})$ in optimal time $\tau = O(k)$. These results are obtained numerically in Section \ref{sub:prob} and confirmed again analytically in Appendix \ref{sec:IntegralMethod}. This leads to an estimated run-time for the model of $ O(k^{5/3})$. The run-time for Feynman's model is a significant improvement over adiabatic evolution O($k^4$) revealing the computational power of this simpler model. In our study, we run into difficulties with the stopping time $\tau$ as it is very precisely defined, after which there are rapid oscillations of the probability amplitude. In Section \ref{sub:oscill}, we show numerically that the time difference between the first and the second maximum behaves like $k^{1/3}$. This is further proven analytically in Appendix \ref{sec:Predicting the Location of the Second Maximum}. This is indeed much shorter than the stopping time, but nevertheless increases with $k$. This is promising for experimental implementations and at least shows for large number of operations $k$, it is possible to capture the program counter at the optimal time. A further discussion into implementations is included in Section \ref{sec:conclusion}.

\section{\label{sec:level4}Run-Time of Feynman's Quantum Computer \protect\\ }

\subsection{Reduced Hamiltonian Method}\label{sub:ham}
The Hamiltonian $\hat{H}$ in principle acts on a $2^{k+1} \times 2^{k+1}$-dimensional Hilbert space. However, the clock part of the space decomposes into independent subspaces, as our Hamiltonian conserves electron number. For example, if the initial clock state is $ \ket{0}_c = |10\cdots 0 \rangle_c$, corresponding to an electron occupying the zeroth site with every other site unoccupied, all subsequent states in the time evolution of the computation process will also have one occupied site.

Restricting the Hamiltonian to this subspace, it acquires the form, 
\begin{equation} \label{eq:H matrix}
\hat{H} =\begin{bmatrix}
0 & \hat{U}_1^\dagger & & & & \\
\hat{U}_1 & 0 & \hat{U}_2^\dagger & & & \\
& \hat{U}_2 & 0 & \hat{U}_3^\dagger & & \\
& & \ddots & \ddots & \ddots & \\
& & & \hat{U}_{k-1} & 0 & \hat{U}_k^\dagger \\
& & & & \hat{U}_k & 0 \\
\end{bmatrix}.
\end{equation}

This reduction is illustrated in the case $k=2$ in Appendix \ref{sec:k=2}.
For example, $q_1^\dagger q_0 |100 \rangle_c = |010 \rangle_c$, which leads 
to the $1,0$-element $\hat{U}_1$ of the above matrix.

The resulting time evolution matrix $\hat{G}(t)=e^{-i\hat{H}t}$ has 
a corresponding structure given in \eqref{eq:G_structure}. 
This can be proven by induction as shown in Appendix~B.
\begin{equation} \label{eq:G_structure} \hat{G}(t)_{ij} = \begin{cases} a_{ij}(t) \hat{U}_i^\dag \cdots \hat{U}_{j-1}^\dag &\text{if $0 \leq i < j \leq k$;} \\
a_{ii}(t) \mathbb{I} &\text{if $0 \leq i=j \leq k$}; \\ 
a_{ij}(t) \hat{U}_{i-1} \cdots \hat{U}_j &\text{if $0 \leq j < i \leq k$.} 
                                      \end{cases}
\end{equation}

The coefficients $a_{ij}(t)$ are the matrix elements of the clock evolution $a_{ij}(t) = (e^{-i \hat{H}_{\rm clock} t})_{ij}$ where, 
\begin{equation}\label{eq:Heff}
\hat{H}_\text{clock}=\begin{bmatrix}
0 & 1 & & & & \\
1 & 0 & 1 & & & \\
& 1 & 0 & 1 & & \\
& & \ddots & \ddots & \ddots & \\
& & & 1 & 0 & 1 \\
& & & & 1 & 0 \\
\end{bmatrix}.
\end{equation}

The reduction of our Hamiltonian to this form highlights the equivalence between the propogation of Feynman's program counter and the propogation of tight-binding electrons or spin waves in a one-dimensional chain. We are only concerned with the $\hat{G}_{k,0}$ element, as it contains the full list of $\hat{U}_j$'s acting in the required order to obtain the desired computation $\ket{\psi_{\rm out}}$.

\subsection{Probability of Computation Completion \& Optimal Time}\label{sub:prob}

The coefficient $a_{k,0}(t)$ is all we need to find $P_k(t)$,
\begin{equation}\label{eq:prob}
    P_k(t) = |\hat{G}(t)_{k0}|^2 = |a_{k0}(t)|^2 = |\bra{k}_ce^{-i\hat{H}_{\rm clock}t}\ket{0}_c|^2,
\end{equation}
which is the desired probability corresponding to the time evolution of the clock space from the initial state $\ket{0}_c$ to the final state $\ket{k}_c$. $\hat{H}_{\rm clock} $ given by equation \eqref{eq:Heff} contains all information required to obtain the probability amplitude $P_k(\tau)$.

The optimal time, $\tau$, can be found by maximising $|a_{k0}(t)|$. $P_k(t=\tau)$ thus represents the maximal probability that the computation is complete and therefore all operations have acted on the register. As shown, each $\hat{U}_j$ matrix encoded in the Hamiltonian described in \eqref{eq:Hamiltonian} can be ignored as far as the determination of the optimal stopping time is concerned.  We must now diagonalise $\hat{H}_{\rm clock}$ just as was done for the adiabatic Hamiltonian $\hat{H}_{\rm clock}(s)$ in (\ref{adiabatic}) to determine the run-time for the model.

\begin{figure}[t]
    \centering
    \includegraphics[width=0.95\linewidth]{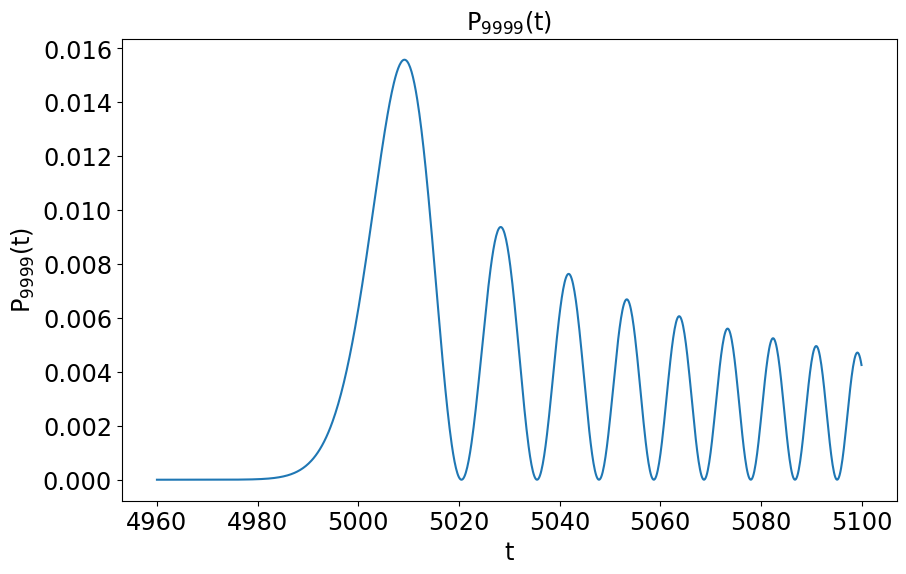}
    \caption{Plot of $P_k(t)$ for k=9999 operators generated from \eqref{eq:summation}. The first local maximum of $P_k(t)$ is also the global maximum and can be represented by $P_k(\tau)$. The amplitude of each local maxima decreases with time if the program is not immediately stopped at $t=\tau$.}
    \label{fig:k=9999}
\end{figure}

This matrix obeys the following eigenvalue equation,

\begin{align*}
    \hat{H}_{\rm clock}\ket{\varphi_j} &= \lambda_j\ket{\varphi_j},\\
  e^{-i\hat{H}_{\rm clock}t}\ket{\varphi_j} &= e^{-i\lambda_jt}\ket{\varphi_j},
\end{align*}

for $j=0,\dots,k$.\\

Using a Fourier transformation, the eigenvalues and eigenstates of the effective Hamiltonian are found to be,
\begin{align}\label{eq:eig}
    \lambda_j=2\cos{\frac{\pi (j+1)}{k+1}}; && \ket{\varphi_j}=\frac{2}{k+1}\begin{bmatrix}
    \sin{\frac{\pi (j+1)}{k+1}} \\
    \sin{\frac{2 \pi (j+1)}{k+1}} \\
    \vdots \\
    \sin{\frac{k \pi (j+1)}{k+1}}
    \end{bmatrix}.
\end{align}
The probability $P_k(t)$ can now be calculated using \eqref{eq:prob},
\begin{equation}\label{eq:summation}
    P_k(t)= \left|\;\frac{2}{k+1}\sum_{j=0}^{k} e^{-i\lambda_jt} \sin^2{\frac{\pi (j+1)}{k+1}}(-1)^{j}\; \right|^2.
\end{equation}

\begin{figure}[t]
\centering
  \includegraphics[width=0.9\linewidth]{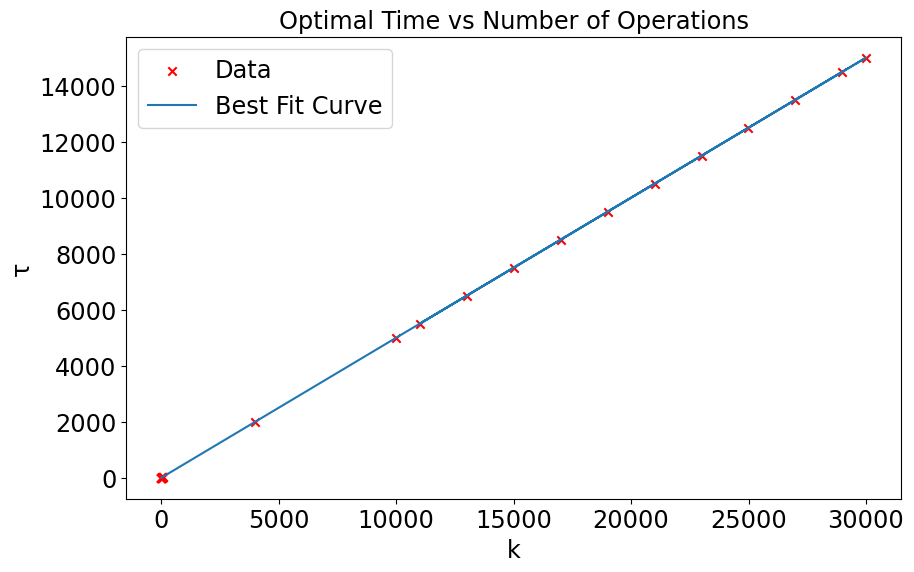}
  \caption{Scatter plot of the optimal time $\tau$ which is found by maximising $P_k(t)$ for a given number of operations $k$. A linear relationship is observed between $\tau$ and $k$ with best fit parameters $\tau = 0.50 k + 2.37$.}
  \label{fig:timevalues}
\end{figure}

Using \eqref{eq:summation} for a large number of operations, $k>>1$, the optimal stopping time $\tau$ is found to be at the first local maximum of $P_k(t)$ as shown in Fig.\ref{fig:k=9999}. There is a small window to capture the first maximal peak after which there are rapid oscillations of probability which decrease in amplitude with time. Numerically, the optimal stopping time is found to grow linearly with  the number of operations, $\tau \propto 0.50 k$,  as shown in Fig.\ref{fig:timevalues}.

\begin{figure}[b]
    \centering \includegraphics[width=0.95\linewidth]{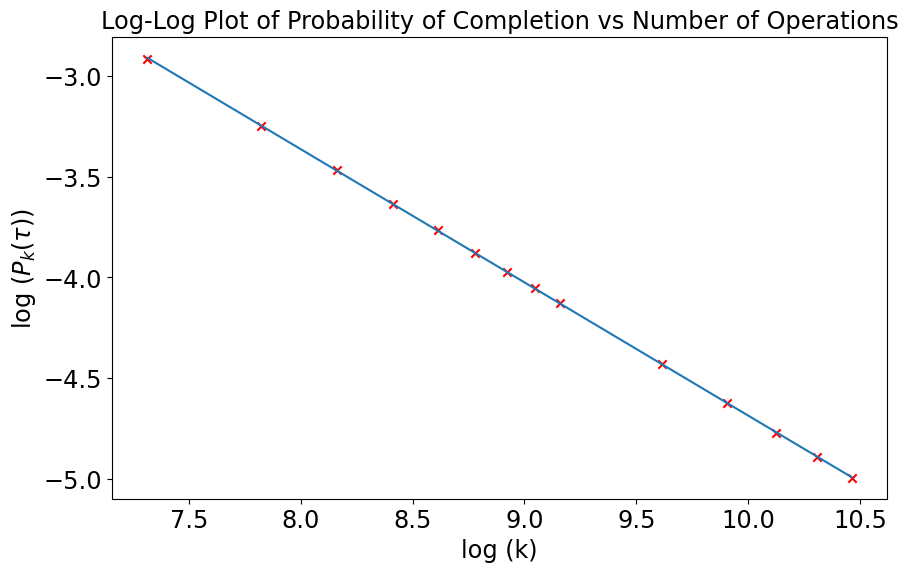}
    \caption{Log-Log scatter plot of the optimal probability $P_k(\tau)$ which is found by maximising $P_k(t)$ for a given number of operations $k$. An inverse cubic relationship is observed between $P_k(\tau)$ and $k$ with best fit parameters $P_k(\tau) = 6.76 k^{-2/3}$.}
    \label{fig:pvsK}
\end{figure}

The probability $P_k(\tau)$ can be plotted against $k$ to derive the relationship between probability of computation completion at the optimal time and the number of operations. Numerically, we find that there is an inverse cubic relationship between them, $P_k(\tau) = \alpha k^{-2/3}$, as shown in Fig.\ref{fig:pvsK}. To obtain a success probability of order 1 for our $P_k(\tau)$, one must repeat the calculation $O(k^{2/3})$ times, leading to an estimated run-time of $O(k^{5/3})$. This is a significant improvement over adiabatic evolution, which has a running time of $O(k^4)$, as analysed by Aharonov et. al. \cite{aharonov2005adiabatic} (see also \cite{Dooley_2020}).
Analytically verifying the relationship with $P_k(\tau)$ and $k$, we use Taylor expansions of trigonometric functions to express \eqref{eq:summation} in the form,
\begin{multline}\label{eq:Pktausum}
    P_k(\tau)  = \left| \;\frac{4}{k+1}\sum_{p=0}^{\frac{k}{2}} \cos^2\left(\frac{\pi(2p-1)}{2(k+1)}\right)  \right. \\
       \left. \times \; \cos \left(\frac{\pi^3}{6(k+1)^2}\left(p-\frac{1}{2}\right)^3 \right) \; \right| ^2.
\end{multline}
This summation form can be approximated by asymptotic analysis, writing it in the form of a Riemann integral  assuming that we have a large number of operations, $k>>1$: see Appendix \ref{sec:IntegralMethod}. The result is,

\begin{equation}\label{Pktauanalytic}
 P_k(\tau) \approx 5.14 k^{-2/3}.
\end{equation}

\begin{figure}[t]
    \centering
    \includegraphics[width=1\linewidth]{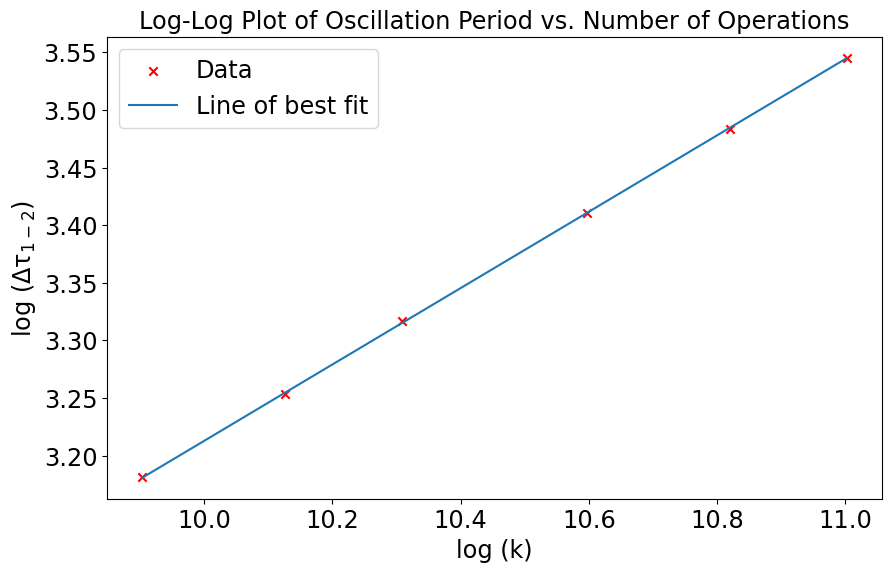}
    \caption{Log-Log plot of time difference, $\Delta \tau_{1-2}$, between the optimal stopping time, $\tau_1$, and the next optimal time $\tau_2$ against number of operations, k.}
    \label{fig:timediff}
\end{figure}

\subsection{Period of Oscillations}\label{sub:oscill}

An initial concern relates to the small window to capture the first maximal peak in Fig. \ref{fig:k=9999} which is followed by large oscillations in decreasing amplitude. It is important to investigate the period of the oscillations after the first maximum of $P_k(t)$, as this indicates how accurately one must be able to determine the stopping time $\tau$ of the computer (to measure the output state). If the register is not stopped at the optimal time, $\tau_1$, but is instead stopped at the next optimal time, $\tau_2$, then the probability $P_k(\tau_2)$ is lower, i.e. it is less likely that the operations have been completed. If $\tau_1 \approx \frac{1}{2}(k+2)$ is the time where the first maximum is attained, we write
$\tau_2\approx\frac{1}{2}(k+2)(1+\delta)$, and $\Delta \tau_{1-2} \approx \delta \times \frac{k+2}{2}$.
Numerically plotting the time difference between the first and second maximum as seen in Fig.\ref{fig:timediff} we find that,
\begin{equation}
    \Delta \tau_{1-2} = \beta k^{1/3}.
\end{equation}
In Appendix \ref{sec:Predicting the Location of the Second Maximum} we obtain the 
analytical prediction  $\Delta \tau_{1-2}=1.115(k+2)^{1/3}$. This means that one needs to be able to stop the computer at a time accurate up to a fraction of $k^{1/3}$. This is much shorter than the stopping time but is still a promising scaling as it increases with $k$. As the number of required operations $k$ grows, the time difference between the first two local maxima also grows, ensuring the optimal time is easier to capture. This also allows the opportunity to capture the register at the second local maximum, removing the need to restart the computation if the optimal stopping time is missed. Assuming that the
accuracy with which the time of measurement is determined is independent of $k$, the scaling found here is optimistic for future implementations.\\

\section{Discussion}\label{sec:conclusion}

We have presented the theory behind Feynman's theoretical construction for a quantum computer, wherein a quantum circuit is mapped to a time-independent Hamiltonian by use of a clock space. A formula is established for the probability, $P_k(t)$, that the desired computation is complete at time t for a quantum computer which executes $k$ number of operations. The analysis here is reminiscent of that for state transfer in quantum spin systems \cite{PhysRevLett.92.187902,christandl2005perfect, sougato}. We demonstrate that the computation is complete
with probability $O(k^{-2/3})$ in optimal time $\tau = O(k)$.
For a success probability of order 1, the calculation must be repeated $O(k^{2/3})$ times, leading to an estimated run time of $O(k^{5/3})$. This is a significant
improvement over adiabatic evolution with a run time $O(k^4)$. We deal with an ideal machine and the superiority of the model is dependent on whether this can be implemented in practice. Although Feynman's clock model has been appreciated on a theoretical level for decades, it is only recently receiving necessary experimental attention. The major challenge with implementation stems from the controlled interaction between the clock register and the unitary gates. Previous avenues of implementation for Hamiltonian quantum computing have included multi-particle quantum walks with interactions using ultra cold bosonic atoms \cite{childs2013universal,lahini2018quantum}. Most interesting to us is a two dimensional lattice construction using a time-independent Hamiltonian by Lloyd and Terhal \cite{lloyd2016adiabatic}. It has been recently expanded on by Ciani et al. in \cite{ciani2019hamiltonian} with a proposal for a superconducting transmon qubit implementation, that exploits the railroad switch modification of the clock register introduced by Nagaj \cite{Nagaj_2010,PhysRevA.85.032330} for universal quantum computation.  The implementation used is similar to those in \cite{janzing2007spin, breuckmann2014space, gosset2015universal}. Ciani provides a direct comparison to the multi-clock Lloyd–Terhal construction by investigating a physical implementation of the single clock Feynman model which we explore here. They discuss how to achieve the weaker hopping interactions and strong-attractive cross Kerr coupling necessary between transmons in this scheme. An initial investigation into the error rates related to gate implementation is discussed in \cite{ciani2019hamiltonian}, but a more thorough analysis into fault tolerance and leakage protection for this 2D rail-road implementation can be found in \cite{lloyd2016adiabatic}. Incorporating quantum error correction explicitly into Feynman's model remains an open question that requires further investigation. It is clear these new routes of implementation for Feynman's computer reinforces the importance of further investigation into this efficient model.

\section*{Acknowledgements}

The authors would like to thank Steve Campbell for valuable discussions and feedback.

\bibliography{apssamp}

\section{Appendix}
\appendix

\section{\label{app: adiabatic} Circuit-to-Hamiltonian construction for Adiabatic Evolution }

In the adiabatic method one evolves the time-dependent Hamiltonian of the form,
$$ H(s) = (1-s)H_{\rm init} + s H_{\rm final}, $$ where $s=t/T$ is a rescaled time as in 
the adiabatic theorem and $T$ is a large final time. During the computation $s$ varies from $0$ to $1$. One must ensure that the groundstate of $H_{\rm init}$ encodes the initial state of our computation  $\ket{\psi_{in}}= \ket{0_1 0_2...0_n}$ and that the groundstate of $H_{\rm final}$ has a non-negligible inner product with the answer of our computation $\ket{\psi_{out}}= \hat{U}_k...\hat{U}_2 \hat{U}_1\ket{\psi_{in}}$. The main ingredient of Aharonov's solution \cite{aharonov2005adiabatic} was the incorporation of Feynman's clock space in the following way.\\
The initial Hamiltonian is given by, 
$$ H_{\rm init} = \sum_{i=1}^N |1_i \rangle \langle 1_i| \otimes |0\rangle_c \langle 0|_c + 
I \otimes \sum_{l=1}^k |l\rangle _c \langle l|_c. $$
(Here $|1_i \rangle \langle 1_i|$ is the projection on the state $|1\rangle$ of the $i$-th qubit, and 
acts as the identity on the other qubits.) This Hamiltonian has the required groundstate $\ket{\psi_0} = \ket{0}_c\otimes\ket{\psi_{in}}$. The final Hamiltonian is, 
$$ H_{\rm final} =  H_{\rm Feynman} + \sum_{i=1}^N |1_i \rangle \langle 1_i| \otimes |0\rangle_c \langle 0|_c. $$ $H_{\rm Feynman}$ here represents Feynman's Hamiltonian introduced in Eq. \ref{eq:Hamiltonian}. This Hamiltonian has the required groundstate $\ket{\psi_{\rm hist}} = \frac{1}{\sqrt{k+1}} \sum_{t=0}^{k} \ket{\psi_t} $ which is a superposition of the $k+1$ orthonormal states $\ket{\psi_t} = \ket{t}_c\otimes U_t,...,U_1 \ket{\psi_{in}}$. The effective evolution of the clock is given by the clock Hamiltonian,
$$
H_{\rm clock}(s) _{l,l'} = \begin{cases}
    s/2 & \text{if } l=l'=0, \\ 
    1 & \text{if } l=l' \in\{1,\dots,k-1\}, \\
    1-s/2 & \text{if } l=l'=k, \\ 
    -s/2 & \text{if } |l-l'|=1, \\ 
    0 & \text{otherwise.}
\end{cases}
$$
The gap in the spectrum between the ground state and the first excited state of this clock Hamiltonian 
determines how large $T$ must be to ensure with high probability that the total system remains in the 
ground state, and the eventual evolution is given by $H_{\rm Feynman}$. (In fact $H_{\rm Feynman}$ 
is defined slightly differently which is the reason that the evolution at $s=1$ is not identical to our 
Hamiltonian \ref{eq:H matrix}). The minimum gap occurs at a time $s$ which is slightly less than 1.

\section{\label{sec:k=2} Feynman's Hamiltonian for k = 2 \protect\\ }

To understand the structure of the Hamiltonian and the time evolution operator, we look at the simplest example of having two operations on our register. For k=2, we add to the n atoms in our register a new set of 3 atoms, which we call 'program counter sites'. $q_i$  and $q_i^\dag$ represent the annihilation and creation operators, respectively, for the program sites i = 0,1,2. Our Hamiltonian in matrix form is given by,

\begin{equation}
    \hat{H}_{\rm TOT} = 
    \begin{bmatrix}
        0&0&0&0&0&0&0&0 \\
        0&0&\hat{U}_2&0&0&0&0&0 \\
        0&\hat{U}_2^\dag&0&0&\hat{U}_1&0&0&0 \\
        0&0&0&0&0&\hat{U}_1&0&0 \\
        0&0&\hat{U}_1^\dag&0&0&0&0&0 \\
        0&0&0&\hat{U}_1^\dag&0&0&\hat{U}_2&0 \\
        0&0&0&0&0&\hat{U}_2^\dag&0&0 \\
        0&0&0&0&0&0&0&0 
    \end{bmatrix}.
\end{equation}

This Hamiltonian relates to 2 independent clock spaces. The first is a program counter where at all times only one site is occupied, i.e $| 100 \rangle$, $| 010 \rangle$, $| 001 \rangle$. The second is a program counter where at all times two sites are occupied, i.e 
 $| 011 \rangle$, $| 101 \rangle$, $| 110 \rangle$. The construction of Feynman's Hamiltonian ensures that the number of program sites occupied is a conserved quantity. This allows us to block-diagonalize the Hamiltonian into non-interacting blocks and separate out the part where only one clock site is occupied.

We will suppose that, in the operation of this computer, only one site is
occupied for all time. We call this Hamiltonian $\hat{H}$ and can extract it from our total Hamiltonian,

\begin{equation}\label{H2}
    \hat{H} = 
    \begin{bmatrix}
        0 & \hat{U}_1\dag & 0 \\
        \hat{U}_1 & 0 & \hat{U}_2\dag \\
        0 & \hat{U}_2 & 0 
    \end{bmatrix}.
\end{equation}

This Hamiltonian satisfies the relation,
\begin{equation}
    \hat{H}^3 = 2\hat{H}.
\end{equation}

Using a Taylor expansion, the time evolution operator reduces to,

\begin{equation}
    e^{-i\hat{H}t} = \hat{I}-i\frac{\hat{H}}{\sqrt{2}}\sin{\sqrt{2}t}+\frac{\hat{H}^2}{2}(\cos{\sqrt{2}t} - 1).
\end{equation}

In matrix form this is,

\begin{equation}
e^{-i\hat{H}t} = 
    \begin{bmatrix}
        1+\frac{\cos{\sqrt{2}t}-1}{2} & \frac{i}{\sqrt{2}}\sin{\sqrt{2}t} \hat{U}_1^\dag & \frac{\cos{\sqrt{2}t}-1}{2} \hat{U}_1^\dag \hat{U}_2^\dag \\
        \frac{-i}{\sqrt{2}}\sin{\sqrt{2}t} \hat{U}_1 & \cos{\sqrt{2}t} & \frac{-i}{\sqrt{2}}\sin{\sqrt{2}t} \hat{U}_2^\dag \\
        \frac{\cos{\sqrt{2}t}-1}{2} \hat{U}_2 \hat{U}_1 & \frac{i}{\sqrt{2}}\sin{\sqrt{2}t} \hat{U}_2 &  1+\frac{\cos{\sqrt{2}t}-1}{2}
    \end{bmatrix}.
\end{equation}

This matrix is unitary. When the program counter reaches the final site $| 001 \rangle$ the data register has been multiplied by the operators $\hat{U}_2\hat{U}_1$ as expected. This occurs at $t=\frac{\pi}{\sqrt{2}}$.

\section{\label{sec:Proving the Structure of a Hamiltonian}Proving the Structure of a Hamiltonian \protect\\ }

To prove the structure given in \eqref{eq:G_structure}, we expand the time-evolution matrix,
\begin{equation}
    \hat{G}(t)=\sum_{m=0}^{\infty}\frac{(-it)^m}{m!}\hat{H}^m.
\end{equation}
Hence it is sufficient to prove that all powers of the Hamiltonian also follow the structure of \eqref{eq:G_structure}, which will be done via a proof of induction. The method of extracting $\hat{H}$, wherein only one site of the clock space is occupied, may be extracted from $\hat{H}_{TOT}$ in a similar manner to \eqref{H2}. The first power of the Hamiltonian is then, by construction,
\begin{center}
\begin{equation}
\hat{H}=\begin{bmatrix}
0 & \hat{U}^{\dag}_{1} & & & & \\
\hat{U}_{1} & 0 & \hat{U}^{\dag}_{2} & & & \\
& \hat{U}_{2} & 0 & \hat{U}^{\dag}_{3} & & \\
& & \ddots & \ddots & \ddots & \\
& & & \hat{U}_{k-1} & 0 & \hat{U}^{\dag}_{k} \\
& & & & \hat{U}_{k} & 0 \\
\end{bmatrix},
\end{equation}
\end{center}
which clearly has the structure in \eqref{eq:G_structure}. 
To prove that $\hat{H}^{m+1}$ has this same structure assuming that $\hat{H}^{m}$ does, we approach the problem case by case. Explicitly, each element is given by,
\begin{align*}
    (\hat{H}^{m+1})_{ij}&=(\hat{H}\,\hat{H}^{m})_{ij},\\
    &=\sum_l^{k}\hat{H}_{il}\hat{H}^m_{lj},\\
    &=\hat{H}_{i,i-1}\hat{H}^m_{i-1,j}+\hat{H}_{i,i+1}\hat{H}^m_{i+1,j}.
\end{align*}

For $i=j$, where $b^{(m)}_{ij}$ is the coefficient associated to $\hat{H}_{ij}^{(m)}$, where,

\begin{equation}
 m=1, b^{(1)}_{ij} = \begin{cases} 
 1 &\text{if $|i-j|=1$;} \\ 0 &\text{otherwise,}\end{cases}
\end{equation}

such that,
 
\begin{align*}
    (\hat{H}^{m+1})_{ii}&=b^{(m)}_{i-1,i}\hat{U}_{i-1}\hat{U}^{\dag}_{i-1}
    +b^{(m)}_{i+1,i}\hat{U}^{\dag}_{i}\hat{U}_{i},\\
    &=(b^{(m)}_{i-1,i}+b^{(m)}_{i+1,i})\mathbb{I},
\end{align*}

where we have used the identity $U^{\dag}_i U_i = \mathbb{I}$ to reduce the string length. Similarly, for $0\leq i < j \leq k$, 
\begin{align*}
    (\hat{H}^{m+1})_{ij}&=b^{(m)}_{i-1,j}\hat{U}_{i-1}\hat{U}^{\dag}_{i-1}\hat{U}^{\dag}_i\ldots \hat{U}^{\dag}_{j-1}+b^{(m)}_{i+1,j}\hat{U}^{\dag}_i\ldots \hat{U}^{\dag}_{j-1}\\
    &=(b^{(m)}_{i-1,j}+b^{(m)}_{i+1,j})\hat{U}^{\dag}_i\ldots \hat{U}^{\dag}_{j-1}.
\end{align*}

Likewise, for $0\leq j < i \leq k$,
\begin{align*}
    (\hat{H}^{m+1})_{ij}&=b^{(m)}_{i-1,j}\hat{U}_{i-1}\ldots \hat{U}_j
    +b^{(m)}_{i+1,j}\hat{U}^{\dag}_{i}\hat{U}_{i}U_{i-1}\ldots \hat{U}_j\\
    &=(b^{(m)}_{i-1,j}+b^{(m)}_{i+1,j})\hat{U}_{i-1}\ldots \hat{U}_j.
\end{align*}
Hence each power of the Hamiltonian has the same structure as \eqref{eq:G_structure}, and therefore $\hat{G}(t)$ must also have the same structure. Note that this derivation also proves that the clock Hamiltonian is given by \eqref{eq:Heff}.

\section{\label{sec:IntegralMethod}Analytic solution of $P_k(\tau)$ \protect\\ }

Using $\hat{G}(t)=e^{-i\hat{H}t}$ and the eigenvalues and eigenvectors given in \eqref{eq:eig}, an analytic solution for $P_k(t)$ can be found,

\begin{equation}
    \begin{split}
    P_k(t)&=\left|\bra{k}\hat{G}(t)\ket{0}\right|^2\\
    &= \biggl|\sum_n \sum_j \braket{k}{\psi_n}\bra{\psi_n}e^{-i\hat{H}_{eff}t}\ket{\psi_j}\braket{\psi_j}{0}\biggr|^2 \\
     &= \biggl|\sum_n \sum_j \braket{k}{\psi_n}\bra{\psi_n}e^{-i\lambda_jt}\ket{\psi_j}\braket{\psi_j}{0}\biggr|^2 \\
     &= \biggl| \sum_j e^{-i\lambda_jt} \braket{k}{\psi_j}\braket{\psi_j}{0}\biggr|^2 \\
     &= \biggl|\;\frac{2}{k+1}\sum_{j=0}^{k} e^{-i\lambda_jt} \sin^2{\frac{\pi (j+1)}{k+1}}(-1)^{j}\; \biggr|^2.
     \end{split}
\end{equation}

In the following, we assume that $k$ is an even number. The derivation for odd $k$ is analogous. Starting from the expression for $a_{k0}(t)$ given in \eqref{eq:summation}, we split the sum into two parts,
\begin{equation}
\begin{split}
    a_{k0} & =\frac{2}{k+1}\sum_{j=0}^{\frac{k}{2}} e^{-i\lambda_jt} \sin^2{\frac{\pi (j+1)}{k+1}}(-1)^{j}\\
    & +\frac{2}{k+1}\sum_{j=\frac{k}{2}+1}^{k} e^{-i\lambda_jt} \sin^2{\frac{\pi (j+1)}{k+2}}(-1)^{j}.
\end{split}
\end{equation}

By a change of variable $j=k+1-m$ in the second term,
\begin{equation}
\begin{split}
    a_{k0} =& \frac{2}{k+1}\sum_{m=0}^{\frac{k}{2}} e^{-i\lambda_mt} \sin^2{\frac{\pi (m+1)}{k+1}}(-1)^{m} \\
    & - \frac{2}{k+1}\sum_{m=0}^{\frac{k}{2}} e^{+i\lambda_mt} \sin^2{\frac{\pi (m+1)}{k+1}}(-1)^{k+1-m} \\
     =&  \frac{2}{k+1}\sum_{m=0}^{\frac{k}{2}} (e^{-i\lambda_mt} - e^{i\lambda_mt} ) \sin^2{\frac{\pi (m+1)}{k+1}}(-1)^{m}\\
    =&  \frac{-4i}{k+1}\sum_{m=0}^{\frac{k}{2}} \sin{\lambda_mt} \sin^2{\frac{\pi (m+1)}{k+1}}(-1)^{m}\\
\end{split}
\end{equation}

Substituting in \eqref{eq:eig} the (approximate) optimal time $\tau = \frac{k+1}{2}$,
we get
\begin{align*}
     a_{k0} &= \frac{-4i}{k+1}\sum_{m=0}^{\frac{k}{2}} 
     \sin \left( (k+1) \cos\left(\frac{\pi (m+1)}{k+1}\right) \right) \\
     & \times \sin^2\frac{\pi (m+1)}{k+1}(-1)^{m}
\end{align*}

Using another change of variable, we set $p = \frac{k}{2}-m$,
\begin{equation*}
\begin{split}
  a_{k0} = & \frac{-4i}{k+1}\sum_{p=0}^{\frac{k}{2}} 
 \sin \left((k+1)\sin\left(\frac{\pi(2p-1)}{2(k+1)}\right) \right)\\
 & \times \cos^2(\frac{\pi(2p-1)}{2(k+1)})(-1)^{\frac{k}{2}-p}
 \end{split}
\end{equation*}

To estimate this sum, we do a second-order Taylor expansion of the argument
of the outer sine-function:

\begin{equation} (k+1) \sin\left(\frac{\pi(2p-1)}{2(k+1)}\right) \approx \left(p-\frac{1}{2}\right) \pi 
 - \frac{\pi^3}{6(k+1)^2}\left(p-\frac{1}{2}\right)^3
\end{equation}

Then noting that $\sin\left(\pi\left(p-\frac{1}{2}\right) - a\right) = (-1)^{p-1} \cos(a)$, we obtain

\begin{equation}
\begin{split}
  a_{(k+1)1} = & \frac{4i (-1)^{\frac{k}{2}}}{k+1}\sum_{p=1}^{\frac{k}{2}} 
  \cos^2\left(\frac{\pi(2p-1)}{2(k+1)}\right)(-1)^{\frac{k}{2}-p} \\ 
  & \times \cos\left(\frac{\pi^3}{6(k+1)^2}\left(p-\frac{1}{2}\right)^3\right).
 \end{split}
\end{equation}

This is formula \eqref{eq:Pktausum}. We now note that the second cosine factor is 
rapidly oscillating for large $k$. This means that large $p$-values make a negligible 
contribution. For small values, the first cosine factor is almost equal to 1 and can be omitted.  Introducing the variable $x = \frac{\pi}{(k+2)^{2/3}} \left(p-\frac{1}{2}\right)$, 
we can approximate the sum by a Riemann integral:

\begin{equation}  a_{k0} \approx \frac{4i (-1)^{\frac{k}{2}}}{\pi(k+1)^{1/3}}
 \int_0^\infty \cos\left(\frac{1}{6} x^3\right)\,dx. 
\end{equation}

The integral can be evaluated using integration by parts, and we obtain 

\begin{equation} a_{k0} \approx 2.27 i(-1)^{k/2} k^{-1/3}. \end{equation}

Squaring this yields \eqref{Pktauanalytic}. There is a difference between our theoretical coefficient of 5.14 and the
numerical value of 6.76. This is due to the approximate form in (D5). Our approximation here holds best at large $k$ and coefficients are expected to show better agreement in this regime.
\\
\\

\section{\label{sec:Predicting the Location of the Second Maximum} Predicting the Location of the Second Maximum \protect\\ }

In Appendix \ref{sec:IntegralMethod}, we found that the first maximum occurs approximately at $\tau =\frac{1}{2}(k+1)$. 
At this value for $t$,
\begin{equation}
    2 t \left( \frac{(2p-1) \pi}{2(k+1)} - \frac{1}{6} \left(\frac{\pi(2p-1)}{2(k+1)} \right)^3 \right)\approx \left(p-\frac{1}{2}\right)\pi
\end{equation}
which is an odd multiple of $\pi/2$. The next maximum should occur when

\begin{equation}
2 t \left( \frac{(2p-1) \pi}{2(k+1)} - \frac{1}{6} \left(\frac{\pi(2p-1)}{2(k+1)} \right)^3 \right)\approx \left(p+\frac{1}{2}\right)\pi.
\end{equation}

Setting $t = (k+2)(1+\delta)/2$, this becomes
\begin{equation}
    \delta \frac{2p-1}{2} - \frac{\pi^2 (2p-1)^3}{48 (k+1)^2} \approx 1
\end{equation}

On the other hand, we want this to remain a good approximation if we increase or decrease $p$. This is the case if we set
\begin{equation}
2t \frac{d}{dp} \left( \frac{(2p-1)}{2(k+1)} -\frac{\pi^2}{6} \left(\frac{\left(p-\frac{1}{2}\right)}{k+1} \right)^3 \right) = 1
\end{equation}
i.e increasing $p$ by 1 also increases the left-hand side by 1. Hence,
\begin{equation}
2t \left( \frac{1}{k+1} -\frac{\pi^2}{2} \frac{\left(p-\frac{1}{2}\right)^2}{(k+1)^3} \right) = 1.
\end{equation}
Inserting $2t = (k+1)(1 + \delta)$ we get
\begin{equation}
\delta \approx  \frac{\pi^2}{2} \frac{\left(p-\frac{1}{2}\right)^2}{(k+1)^2}.
\end{equation}
Combining this with the previous identity we obtain
\begin{equation}
\frac{\pi^2}{24} \frac{(2p-1)^3}{(k+1)^2} = 1.
\end{equation}
Finally inserting this into $\delta$, we get
\begin{equation}
\delta = 0.5 (3\pi)^{2/3} (k+1)^{-2/3} = 2.23 (k+1)^{-2/3}.
\end{equation}

\end{document}